\documentstyle[12pt]{article}

\global\arraycolsep=1pt
\newcommand{\beq}{\begin{equation}}
\newcommand{\eeq}{\end{equation}}
\newcommand{\beqa}{\begin{eqnarray}}
\newcommand{\eeqa}{\end{eqnarray}}
\newcommand{\CR}{\nonumber \\}

\newcommand{\trace}{\hbox {Tr}~}
\newcommand{\LG}{{  \cal G}}
\newcommand{\tod}{\stackrel{d}{\longrightarrow}}

\begin{document}

\renewcommand{\thefootnote}{\fnsymbol{footnote}}
\begin{titlepage}
\null
\begin{flushright}
April  1997 \\
hep-th/9705127
\end{flushright}
\vskip 5em
\begin{center}
{\Large \bf
Cohomological Yang-Mills Theory \\
in Eight Dimensions\footnote{
Talk presented by H.K. at APCTP Winter School
on Dualities in String Theory, (Sokcho, Korea),  February 24-28, 1997.}
\par}
\lineskip .75em
\vskip 3em
\normalsize
{\large  Laurent Baulieu} \\
{\it LPTHE, Universit\'es Paris VI - Paris VII, Paris,
France}\\
\vskip 1 em
{\large Hiroaki Kanno} \\
{\it Department of Mathematics, Faculty of Science,
Hiroshima University, \\
Higashi-Hiroshima 739, Japan}\\
\vskip 1 em
and
\vskip 1 em
{\large I. M. Singer} \\
{\it  Department of Mathematics, MIT, Cambridge, USA}
\end{center}
\vskip 1 em

\begin{abstract}
We construct nearly topological Yang-Mills theories on eight dimensional
manifolds with a special holonomy group.
These manifolds are the Joyce manifold with $Spin(7)$ holonomy and
the Calabi-Yau manifold with $SU(4)$ holonomy.
An invariant closed four form $T_{\mu\nu\rho\sigma}$ on the manifold
allows us to define an analogue of the instanton equation, which serves as
a topological gauge fixing condition in BRST formalism.
The model on the Joyce manifold is related to the eight dimensional 
supersymmetric Yang-Mills theory. Topological dimensional reduction
to four dimensions gives non-abelian Seiberg-Witten equation.
\end{abstract}

\end{titlepage}
\renewcommand{\thefootnote}{\arabic{footnote}}
\setcounter{footnote}{0}


\section{Introduction}

Almost a decade ago topological quantum field theory (TQFT) was proposed 
and it was conjectured that TQFT describes \lq\lq topological phase\rq\rq\ 
where general covariance is unbroken \cite{Wi1}.  The topological phase
was supposed to be a key to the underlying principle of string theory.
It has been shown that in two dimensional quantum gauge theory 
and quantum gravity we can see the topological phase realized
by TQFT. Furthermore, TQFT, or more specifically, 
cohomological quantum field theory is 
related to supersymmetric (SUSY) quantum field theory by twisting.
From mathematical viewpoint, TQFT provides a field theoretical description of
intriguing topological invariants
such as Donaldson invariants, Gromov-Witten invariants 
and Seiberg-Witten invariants.  
So far TQFT has been extensively
studied in two, three and four dimensions. (See e.g. \cite{BBRT},
\cite{CMR} and references therein.)

Recent developments in dualities and non-perturbative
dynamics of superstring theory have stimulated
a renewed interest in SUSY
field theories in diverse dimensions, 
which appear in various compactifications or limits
of M-theory, F-theory, or any other hypothetical
non-perturbative formulation of \lq\lq string
theory\rq\rq. 
From such a viewpoint it is an interesting problem
to see how the idea of TQFT is extended to higher dimensions.
In this article we will report a recent progress in this direction \cite{BKS}.
Our main results are summarized as follows;
we have constructed cohomological Yang-Mills theories in eight dimensions.
On the eight dimensional Joyce manifold and 
the Calabi-Yau fourfold, which have special holonomy groups,
we can find a topological action and 
topological gauge fixing conditions in BRST formalism.
Our covariant gauge conditions are the octonionic instanton equation and the
complexified anti-self-dual (ASD) equation, respectively.
The theory on the Joyce manifold is a twist of eight dimensional
SUSY Yang-Mills theory and hence related to ten dimensional $N\!=\!1$ 
SUSY Yang-Mills theory. 
Moreover, an appropriate dimensional reduction to four
dimensions gives topological field theory of
non-abelian Seiberg-Witten (monopole) equation.
For a full account of technical details and further references, 
we refer to the original paper \cite{BKS}.

The most crucial point for our construction in higher
dimensions is to find an appropriate set of topological
action and covariant gauge fixing conditions.
To obtain a quantum field theory which explores a moduli space of
gauge fixing conditions by a well-defined
weak coupling expansion,
all the dynamical fields should have non-degenerate
kinetic terms after eliminating the auxiliary fields.
This means mathematically we should have a complex of
elliptic operators behind our choice of gauge fixing conditions.
We have found that as covariant gauge fixing condition
the following generalization
of the instanton equation works well for pure Yang-Mills theory in higher
dimensions;
\beq
\frac{1}{2} T^{\mu\nu\rho\sigma} F_{\rho\sigma}~=~\lambda F^{\mu\nu}~, 
\label{inst}
\eeq
where $T^{\mu\nu\rho\sigma}$ is a totally
antisymmetric tensor and $\lambda$ is a constant (an eigenvalue).  
In four dimensions we can choose
the $SO(4)$ invariant tensor $\epsilon^{\mu\nu\rho\sigma}$,
which leads to the self-dual $(\lambda =1)$ or 
the anti-self-dual $(\lambda =-1)$ condition for 
the curvature two form $F_{\mu\nu}$.
If the dimension $N$ is  higher than four,
$T^{\mu\nu\rho\sigma}$ cannot be invariant under $SO(N)$ any more.
Corrigan et al \cite{CDFN} classified  possible choices of 
$T^{\mu\nu\rho\sigma}$ up to eight dimensions, 
requiring that $T^{\mu\nu\rho\sigma}$ is
invariant under a maximal subgroup of $SO(N)$.  They found that
a choice of $Spin(7) \subset SO(8)$ gave seven first order conditions 
for the gauge field $A_\mu$ in eight dimensions.
We adopt these conditions as topological gauge conditions for the
Joyce manifold. In the following we will mainly be devoted to this case.
The Calabi-Yau manifold gives another possibility,
where the complexified
ASD instanton equation is employed as covariant gauge fixing condition.

Twisting of SUSY quantum field theory is
another approach to construct a model of TQFT \cite{Wi1} \cite{EY}.
If the base manifold admits a covariantly constant spinor $\zeta$, 
the supersymmetry transformation with the parameter $\zeta$ 
may be identified as a topological BRST transformation.
But this cannot take place on generic curved manifolds.
To circumvent the problem we use the trick of twisting that
effectively produces a covariantly constant spinor parameter in lower
dimensions.
This is regarded as a result of gauging the R-symmetry,
or global symmetry of a supersymmetric theory.
More generally, we have a chance of having 
covariantly constant spinors due to a reduction of the holonomy group.
This possibility of twisting the supersymmetry 
when the holonomy group is reduced
was first pointed out in \cite{GO}
and it was applied to the twisting of $N\!=\!1$ supersymmetry on four 
dimensional K\"ahler manifolds in \cite{Joh}.
Eight dimensional (oriented) Riemannian manifolds have the holonomy
group $SO(8)$. On the Joyce manifold the holonomy is
reduced to a maximal subgroup $Spin(7)$ \cite{Jo1}. This reduction
allows exactly one covariantly constant spinor, in terms of
which we can define topological BRST transformation.
In this sense the cohomological Yang-Mills theory
on the Joyce manifold is a natural generalization
of the topological Yang-Mills theory on $K3$ surface which
has $SU(2) \simeq Spin(3)$ holonomy \cite{VW}.

\section{Instanton Equation on the Joyce manifold}

Four dimensional topological Yang-Mills theory 
can be obtained by the BRST formalism \cite{BS}.
Taking $p_1= \frac{1}{8\pi^2} \trace (F \wedge F)$ as a topological
Lagrangian, we gauge-fix its topological invariance with
three covariant gauge conditions of (anti-) self-duality $F_{\mu\nu}
\pm \frac{1}{2} \epsilon_{\mu\nu\rho\sigma}F^{\rho\sigma}=0$
and one Feynman-Landau
type gauge condition $\partial_\mu A^\mu=0$. 
It is crucial in this construction that
the (anti-) self-duality of the curvature 
gives three conditions for four degrees of freedom
of local deformations of the gauge field $A_\mu$.  
More mathematically,
we have an elliptic complex $0\to \Lambda^0\tod \Lambda^1 \tod
\Lambda^2_\pm \to 0$, tensored with a Lie algebra $\LG$.
Topological Yang-Mills theory is also 
regarded as a twisted $N\!=\!2$ SUSY Yang-Mills
theory in four dimensions.
Hence, we expect that topological Yang-Mills theory probes 
the vacuum structure of supersymmetric gauge theory.

We want to extend this scheme of topological Yang-Mills theory
to eight dimensions. We cannot do it for an arbitrary manifold.
It is possible when the holonomy group $SO(8)$ is reduced to $Spin(7)$
or $SU(4)$, which allows an invariant closed four form $\Omega$.
We can consider a topological action
\beq
S_0~=~\frac{1}{2} \int_{M_8}~ \Omega \wedge 
\trace ( F \wedge F)~, \label{action}
\eeq
which is quadratic in the curvature $F$.
Since $\Omega$ is closed, $S_0$ is invariant 
under an arbitrary local deformation of the gauge field $A_\mu$; 
$\delta A_\mu = \epsilon_\mu$.
A $Spin(7)$ structure on an eight dimensional (Riemannian) manifold $M_8$ is
given by a closed self-dual $Spin(7)$ invariant four form $\Omega$.
Then, the holonomy group is
a subgroup of $Spin(7)$ and the metric $g$ is Ricci-flat.
In this case $(M_8,g)$ is called the Joyce manifold \cite{Jo1}.
For the Calabi-Yau fourfolds with $SU(4)$ holonomy, we can take 
a covariantly constant $(4,0)$ form as $\Omega$. 

On eight dimensional manifolds, the vector, the chiral
spinor and the anti-chiral spinor are all eight dimensional
representations. They will be denoted by ${\bf 8_v, 8_s, 8_c}$,
respectively. The triality operation of $SO(8)$ interchanges
these representations. 
The reduction of the holonomy group to $Spin(7)$ induces
an invariant decomposition of the chiral spinor; 
${\bf 8_s} = {\bf 1} \oplus {\bf 7}$. The representations
${\bf 8_v}$ and ${\bf 8_c}$ remain irreducible under $Spin(7)$.
The singlet in ${\bf 8_s}$ means the existence of 
a covariantly constant spinor $\zeta$. 
Then the following fourth rank antisymmetric
tensor is $Spin(7)$ invariant;
\beq
T^{\mu\nu\rho\sigma}~=~\zeta^T \gamma^{\mu\nu\rho\sigma} \zeta~,
\label{omega}
\eeq
where $\gamma^{\mu\nu\rho\sigma}$ is the totally antisymmetric
product of $\gamma$ matrices for the $SO(8)$ spinor representation.
Eq.(\ref{omega}) gives a component representation of 
the four form $\Omega$.
On the Joyce manifold with
the invariant fourth rank tensor $T^{\mu\nu\rho\sigma}$,
we can define the following generalization
of the instanton equation \cite{CDFN};
\beq
F^{\mu\nu}~=~\frac{1}{2} T^{\mu\nu\rho\sigma} F_{\rho\sigma}~. \label{Jinst}
\eeq
The curvature 2-form $F_{\mu\nu}$ in 8 dimensions has 28 components,
whose $Spin(7)$ decomposition is 
${\bf 28 = 21 \oplus 7}$. This is made explicit
by the eigenspace decomposition of the action of 
$1/2~T^{\mu\nu\rho\sigma}$ with the eigenvalues
$\lambda = 1$ and $\lambda = -3$. 
Eq. (\ref{Jinst}) means the curvature has
no components in the latter subspace which is 7-dimensional.
Computed explicitly, eq. (\ref{Jinst}) leads the following
seven linear relations among the curvature;
\beqa
\Phi_1 &\equiv& F_{12} + F_{34} + F_{56} + F_{78} = 0~, \CR
\Phi_2 &\equiv& F_{13} + F_{42} + F_{57} + F_{86} = 0~, \CR
\Phi_3 &\equiv& F_{14} + F_{23} + F_{76} + F_{85} = 0~, \CR
\Phi_4 &\equiv& F_{15} + F_{62} + F_{73} + F_{48} = 0~,  \label{gaugefix} \\
\Phi_5 &\equiv& F_{16} + F_{25} + F_{38} + F_{47} = 0~, \CR
\Phi_6 &\equiv& F_{17} + F_{82} + F_{35} + F_{64} = 0~, \CR
\Phi_7 &\equiv& F_{18} + F_{27} + F_{63} + F_{54} = 0~. 
\nonumber 
\eeqa
Behind these combinations there is the algebra of octonions \cite{GG}.
In fact eq.(\ref{gaugefix}) can be written as;
\beq
F_{8i}~=~c_{ijk}~F_{jk}~,  \quad (1 \leq i,j,k \leq 7)
\eeq
where $c_{ijk}$ are the structure constants for
octonions.  For this reason, we refer to eq.(\ref{gaugefix}) as
the octonionic instanton equation.
It is worth while to compare it with the anti-self-dual (ASD) equation;
\beq
F_{4i}~=~\epsilon_{ijk}~F_{jk}~,  \quad (1 \leq i,j,k \leq 3)
\eeq
where we have $\epsilon_{ijk}$; the structure constants 
for quaternions.
On the Calabi-Yau fourfold we can introduce four complex coordinates
and the complexified ASD equation makes sense.
We employ it as topological gauge fixing condition that gives
six conditions on the gauge field.

It is straightforward to check the identity;
\beq
4 \sum_{i=1}^7 \trace (\Phi_i \Phi_i)\cdot~(vol)~=~- \Omega \wedge 
\trace (F \wedge F) + \trace (F \wedge * F)~, \label{square}
\eeq
where $(vol)$ stands for the volume form.
The first term in the righthand side is a density of 
a topological invariant, since $\Omega$ is closed.
The second term is nothing but the standard action
density for the Yang-Mills theory.
Hence a solution to $\Phi_i=0,~(i=1,\ldots, 7)$ gives
a stationary point of the eight dimensional Yang-Mills
theory. In this sense, eq. (\ref{gaugefix}) deserves to 
be called instanton equation on the Joyce manifold.
It is known that the octonionic instanton exists, 
if the gauge group has a $Spin(7)$ subgroup.
The construction is a simple generalization of
the one for BPST instanton in four dimensions, which
exists if the gauge group contains an $SU(2) \simeq Spin(3)$ factor.

Once we find the octonionic instanton equation which serves
as seven covariant gauge conditions, it is straightforward
to perform BRST gauge fixing of the topological action (\ref{action}).
Precisely speaking, however, the model is not a topological quantum field
theory,
since it involves a choice of the four form $\Omega$.
It is topological only in the gauge sector and
the gravitation is treated as a background.
The topological BRST transformations and the final action
are in parallel to those of four dimensional case.
Hence, we omit their expressions, refering to \cite{BKS}.
The mathematics of the moduli
space of the octonionic instanton equation is still an
interesting open problem.


\section{Link to Supersymmetry}

The field contents of the theory on the Joyce manifold are 
in parallel to those of four dimensional topological
Yang-Mills theory.  We have a gauge field $A_\mu$ with the
topological ghost $\psi_\mu$ and the secondary ghost $\phi$.
According to our choice of the seven gauge fixing conditions,
we introduce the anti-ghost $\chi_i~(1\leq i \leq 7)$.  
Finally there also exists a pair of fields $(\eta, \lambda)$
for the gauge fixing of the topological ghost $\psi_\mu$.
Four dimensional topological Yang-Mills theory is a twisted version 
of $N\!=\!2$ super Yang-Mills theory. 
Moreover, it is related by dimensional
reduction to the minimal six dimensional super Yang-Mills theory.
The above field contents naturally lead 
a similar connection in eight dimensions.
In fact this is understood as follows;
the gauge supermultiplet in eight dimensions consists of one
gauge field in ${\bf 8_v}$, one chiral spinor in ${\bf 8_s}$, 
one anti-chiral spinor in ${\bf 8_c}$ and two scalars \cite{SS}.
The reduction of the holonomy group to $Spin(7)$ defines
a well-defined decomposition of the chiral spinor;
${\bf 8_s} = {\bf 1} \oplus {\bf 7}$.  Now it is natural to identify
$A_\mu$ and $\psi_\mu$ in our topological theory 
as ${\bf 8_v}$ and ${\bf 8_c}$, respectively.
Furthermore $\chi_i$ and $\eta$ just correspond to the chiral spinor
${\bf 8_s}$ according to the above decomposition. 
Finally $\phi$ and $\lambda$ give the
remaining two scalars. This exhausts all the dynamical fields in
our action of eight dimensional cohomological Yang-Mills theory.
Thus, the theory on the Joyce manifold is identified as a twisted
super Yang-Mills theory in eight dimensions.
One can show that the supersymmetry transformation with a covariantly
constant spinor as its parameter recovers the BRST transformation.

The supersymmetric Yang-Mills theory in eight dimensions
is obtained by dimensional reduction 
from the ten dimensional $N\!=\!1$ super Yang-Mills theory.
It has been argued that the effective world volume theory of
the D-brane is the dimensional reduction of the ten dimensional
super Yang-Mills theory \cite{Wi2}.  Hence the cohomological Yang-Mills
theory constructed in this paper may arise as an effective
action of 7-brane theory.  
We expect that the world volume theory
of D-branes would provide a variety of higher dimensional
TQFT's \cite{BSV}.


\section{Topological Dimensional Reduction}

It is amusing that by dimensional reduction 
the octonionic instanton equation (\ref{gaugefix}) gives 
other intriguing equations in lower dimensions.
For example, a dimensional reduction to six dimensions decomposes
eq.(\ref{gaugefix}) into two sets; a single condition 
$\Phi_1=0$ and others $\Phi_k=0,
~(2 \leq k \leq 7)$. These are equivalent to
the Donaldson-Uhlenbeck-Yau equation for the moduli problem of
stable holomorphic vector bundles.
More interestingly
an appropriate topological dimensional reduction to four dimensions
recovers a non-abelian version of Seiberg-Witten (monopole) equation.
In this case the octonionic instanton equation can be separated into 
3 plus 4 equations. 
We group $A_5, A_6,A_7,A_8$ into a four component field
$\varphi^\alpha$, $\alpha =1,2,3,4$, which can be interpreted 
as a commuting complex Weyl spinor on a Cayley 
submanifold $C_4$ of the Joyce manifold \cite{BSV}.
That is, the normal bundle to $C_4$ is isomorphic to 
(twisted) spin bundle on $C_4$.
Of course, $A_\mu=A_1,
A_2,A_3,A_4$ are gauge fields on $C_4$.
Then the first three of eq.(\ref{gaugefix}) imply
\beq
F_{\mu\nu}+\epsilon_{\mu\nu\rho\sigma }
F^{\rho\sigma} + \varphi^T \Sigma _{\mu\nu}\varphi=0~, \label{monopole1}
\eeq
where $\Sigma_{\mu\nu}= \frac{1}{4} [ \Gamma_\mu, \Gamma_\nu]$ is 
the anti-symmetric product of the gamma matrices in four dimensions.
The remaining four equations are written as a Dirac-type equation
\beq
D^{(A)}_\mu \Gamma^\mu \varphi=0~.  \label{monopole2}
\eeq
We recognize eqs.(\ref{monopole1}) and (\ref{monopole2}) 
as non-abelian Seiberg-Witten equation 
with a matter $\varphi$ in the adjoint representation.
Actually this is not unexpected, if we recall the link to SUSY 
discussed in the last section, where
the octonionic instanton equation in 8 dimensions is connected
the dimensional reduction of
the $N\!=\!1$ super Yang-Mills theory in 10 dimensions. 
Therefore, we predict that the theory we 
obtain by dimensional reduction to 4
dimensions is related to a twisted version of the $N\!=\!4$ super Yang-Mills
theory \cite{VW}. 

We thus conclude that
the fields of the eight dimensional cohomological Yang-Mills theory,  
the fields which are involved in the four dimensional Seiberg-Witten
equations, and the fields  of the $D\!=\!4, N\!=\!4$ super Yang-Mills
theory are all connected by twist and dimensional reduction.
The ten dimensional $N\!=\!1$ super Yang-Mills theory underlies all these
theories.

\section{Conclusion}

We have shown that we can construct cohomological Yang-Mills theories in eight
dimensions. The reduction of the holomony group to $Spin(7)$ or $SU(4)$ allows
an invariant closed four from $\Omega$, which we have used for both
topological action and covariant gauge fixing condition.
A comparison of two cases is made in the Table 1.
We expect that a model on the eight dimensional hyperK\"ahler manifold
with $Sp(2) \simeq Spin(5)$ holonomy is also interesting.

\begin{table}
\caption{8 dimensional Cohomological Yang-Mills Theories}
\begin{center}
\begin{tabular}{|c||c|c|}
\null & Joyce & Calabi-Yau  \\ 
\hline
holomomy & $Spin(7)$ &  $SU(4) \sim Spin(6)$  \\
\hline
closed 4 form & $T^{\mu\nu\rho\sigma} = 
\zeta^T \gamma^{\mu\nu\rho\sigma}\zeta$ & 
 holomorphic (4,0) form $\Omega$  \\
\hline
division algebra & {\bf O}   &   {\bf C}$\times${\bf H}   \\
\hline
topological gauge  & octonionic instanton & 
complexified ASD \\
fixing  & 7 conditions & 6 conditions \\
\hline
ordinary gauge fixing & real Lorentz condition & complex Lorentz condition \\
\hline
topological BRST &  $N_T = 1$  &  $N_T = 2$
\end{tabular}
\end{center}
\end{table}

Going to the higher dimensions gives other types of topological
models involving higher rank gauge fields. In fact we have constructed
several models up to 12 dimensions. For more details, see \cite{BKS}.

\vspace{5ex}

{\bf Acknowledgements;}

H.K. would like to thank the organizers of the workshop
for providing an opportunity to announce our work.
He is also grateful to local organizers, especially to Soonkeon Nam,
for their kind hospitality. 
This work is supported in part
by the Grant-in-Aid for Scientific Research 
from the Ministry of Education, Science and Culture, Japan.

\bigskip


\end{document}